# Collision-sticking kinetics of acid-base clusters and its influence on atmospheric new particle formation

Huan Yang*[1], Bernhard Reischl[1], Ivo Neefjes[1], Valtteri Tikkanen[1], Jakub Kubečka[1], Theo Kurtén[2], and Hanna Vehkamäki[1]

[1]Institute for Atmospheric and Earth System Research/Physics, University of Helsinki, FI-00014 Helsinki, Finland;
[2]Institute for Atmospheric and Earth System Research/Chemistry, University of Helsinki, FI-00014 Helsinki, Finland;

*Corresponding Author: Huan Yang, **Email:** huan.yang@helsinki.fi

**Abstract**

Kinetics of collision-sticking processes between vapor molecules and molecular clusters of low volatile compounds facilitates the initial steps of atmospheric clustering. Conventional theoretical models are quite inaccurate due to the neglect of long-range interactions that essentially govern the kinetics of these microscopic phenomena. Here, we present a consistent and generic theoretical model for evaluating collision rates between molecules and molecular clusters with intermolecular potentials properly incorporated. The model requires solely the elementary molecule-molecule potential as a-priori information but predicts collision rates of molecular clusters at arbitrary sizes, with an accuracy comparable to all-atom molecular dynamics simulations we performed for sulfuric acid-dimethylamine clusters, a typical example of acid-base induced clustering. The carefully devised simulations validate the theoretical model and elucidate the kinetics of the molecular collision-sticking process. It is found that the vibrational coupling after the collision between a sulfuric acid molecule and a sulfuric acid cluster can be occasionally unsuccessful, i.e., no stable bond is formed after the collision. However, introducing dimethylamine molecules to the sulfuric acid cluster can notably increase the probability of forming stable bonds and hence stable product clusters. The results offer fundamental insights into the initial steps of molecular clustering and will facilitate the development of efficient kinetic approach-based nucleation models.

**Introduction**

Formation of condensed-phase clusters (1) from super-saturated vapors is the elementary step for atmospheric new particle formation (2-6). Precise quantifications of the cluster formation rates are important in assessing global aerosol budgets and their subsequent climate (7-9) and health effects (10, 11). The first steps of cluster formation can be thought of as a microscopic kinetic process where the size evolution of new-born clusters is facilitated by competitive events of monomer (12) acquisition and loss; corresponding theoretical models resort to solving a set of pseudo-first-order kinetic equations (13):

$$\frac{dN_i}{dt} = k_{i-1,1}N_{i-1}N_1 - k_{i,1}N_iN_1 - \gamma_i N_i + \gamma_{i+1}N_{i+1}, \qquad i = 1, 2, \ldots, \qquad (1)$$

where $N_i$ denotes the number concentration of clusters composed of $i$ monomers, $k_{i,1}$ is the monomer-cluster collision rate coefficient, and $\gamma_i$ is the evaporation rate coefficient. Eq. **1** (here written, for simplicity, for a one component system) is the basis for numerous theoretical nucleation models (14, 15) including the commonly used classical nucleation theory (16, 17) (CNT). Although the criterion that the monomer condensation and evaporation dominate over the cluster-cluster collision and fragmentation events applies to many situations (18), it is still found that CNT agrees with experiments for only a narrow range of saturation ratios and temperatures. Various extensions and revisions to CNT (19-25) have been made, but a precise and generic model is yet to be developed. In CNT, collision rates are derived from kinetic gas theory where intermolecular potentials are neglected while evaporation rates are calculated through a detailed balance against condensation under equilibrium conditions (26); these approximations may lead to significant errors in the initial steps of clustering. First, the evaluation of the equilibrium constants involves bulk material properties that may not accurately describe the behavior of the smallest clusters, introducing large inaccuracies in the evaporation rates. Further, the kinetic gas theory may not predict accurate collision rates, as intermolecular attractive long-range forces can enhance mutual collisions, especially for atmospheric clusters composed of polar molecules. This will lead to additional inaccuracies in the evaporation rates. Moreover, using collision rates to approximate monomer uptake rates may be questionable especially for the smallest clusters, as the formation of stable bonds (and hence stable product clusters) between the monomer and cluster after a collision may be unsuccessful in a significant fraction of the events, i.e., the mass accommodation coefficient or sticking probability can be much below unit. Recently, we have developed molecular dynamics frameworks (27, 28) that predict enhanced collision rates in good agreement with experiments (29, 30). Molecular dynamics approaches directly monitor the trajectories of all atoms under empirical force fields, accounting for the consequences of all possible molecular interactions. However, as adequate samplings of space and velocity distributions leads to high computational cost, the prospects for generalizing these approaches are limited. Accurate methods that can be efficiently applied to a wide variety of clusters and molecules remain to be proposed.

Atmospheric aerosol formation is essentially a process described by Eq. **1** with low volatile trace compounds as the clustering molecules. However, the concentrations of participating vapors, e.g., sulfuric acid, and one-component cluster stabilities inferred from state-of-the-art quantum chemical calculations cannot explain formation rates observed in field measurements. Th enhanced rates are thought to be achieved by the stabilization effects of a mixture of molecules participating in the initial steps of clustering (31). For example, dimethylamine $[(CH_3)_2NH]$ has shown to be an effective base to stabilize sulfuric acid $(H_2SO_4)$ clusters through a proton transfer reaction:

$$H_2SO_4 + (CH_3)_2NH \rightleftharpoons (CH_3)_2NH \cdot H_2SO_4 \rightarrow (CH_3)_2NH_2^+ \cdot HSO_4^-. \qquad (2)$$

Under atmospheric conditions, the product cluster $(CH_3)_2NH_2^+ \cdot HSO_4^-$ is much less volatile compared to reactants, and hence the cluster formation pathway involves the sequential addition of sulfuric acid and dimethylamine molecules (32, 33). Chamber experiments have shown that, for the sulfuric acid-water system, dimer and trimer formation can be the bottleneck of nucleation: critical clusters may contain only two or three sulfuric acid molecules (34, 35). The presence of base molecules even at part-per-trillion mixing ratios can potentially eliminate the barrier, making cluster formation approach the kinetic limit with minimal evaporation. In these cases of barrierless new particle formation, the last two terms in Eq. **1** can be dropped without affecting the model precision, which further emphasizes the importance of accurately predicting cluster collision rates in achieving a generally applicable model for atmospheric clustering.

**Results and Discussion**

**Binary collisions**. We consider a general starting point of a binary collision where two colliding entities, treated as point particles, are originally located at $\vec{r}_{oi}$ and $\vec{r}_{oj}$ with respective initial velocities of $\vec{v}_i$ and $\vec{v}_j$. They start with an impact parameter of $b$ defined as the magnitude of the cross product of their relative position vector and a unit vector aligned with their relative velocity, i.e., $b = |(\vec{r}_{oi} - \vec{r}_{oj}) \times (\vec{v}_i - \vec{v}_j)|/|\vec{v}_i - \vec{v}_j|$ (schematics shown in Fig. 1B). The critical impact parameter $b_c(v)$ is the maximum value of $b$ that can result in a collision for a given relative speed $v = |\vec{v}_i - \vec{v}_j|$. The attractive potential between the colliding entities leads to $b_c(v) > R_i + R_j$, where $R_i + R_j$ is the sum of entities' hard-sphere radii; note that the collision is defined as the event where the center of mass of the two entities are, for the first time, within a "distance of influence" taken as $R_i + R_j$. In dilute gaseous systems, the binary collision rate coefficient can be derived by considering the flux of one particle onto the other, impinging from a circular region with a radius of $b_c$, which is "infinitely far" and orientated to an arbitrary angle (36):

$$k_{i,j} = \pi \int_0^\infty dv \; b_c^2(v) \, v \, f(v), \tag{3}$$

where $f(v)$ is the Maxwell-Boltzmann distribution of the relative speed. The problem is then reduced to evaluating $b_c$. To that end, we resort to an hypothetical "minimum distance" $r_m$, the minimum center-to-center distance the two interacting entities can reach when orbiting around each other at relative speed $v_m$, that is linked to the initial impact condition (i.e., a given $b$ and $v$ pair) through conservation of energy and angular momentum (37):

$$\tfrac{1}{2} m_{ij} v^2 = \tfrac{1}{2} m_{ij} v_m^2 + U(r_m), \tag{4a}$$

$$m_{ij} v \, b = m_{ij} v_m \, r_m, \tag{4b}$$

leading to:

$$b^2 = r_m^2 \left[ 1 - \frac{2 U(r_m)}{m_{ij} v^2} \right] \equiv \omega_v(r_m), \tag{5}$$

where $m_{ij} = m_i m_j / (m_i + m_j)$ is the reduced mass, $U(r_m)$ is the potential energy of the two interacting entities at distance $r_m$, and the function $\omega_v(r_m)$ is defined by Eq. **5** for convenience purposes. Note that the interaction potential $U$ at the initial state is neglected in Eq. **4a** and the relative velocity $\vec{v}_m$ is perpendicular to the center-to-center line for the minimum distance $r_m$. For common functional forms of attractive potential $U(r_m) \propto -r_m^{-n}$, in the interval of $r_m \geq 0$, function $\omega_v(r_m)$ is either a concave function ($n \geq 3$) exhibiting a single minimum at a distance denoted by $R_m$, descending when $0 \leq r_m < R_m$ and increasing when $r_m \geq R_m$, or a monotonically increasing function ($n = 1$ or $2$) with the minimum located at $R_m = 0$. Eq. 5 serves as a criterion for determining whether a collision can happen: if Eq. 5 does not have any real positive roots $r_m$ for a given set of impact condition $(b, v)$ and intermolecular potential $U$, then collision is ensured, as this corresponds to the physical scenario where one entity infinitely approaches the center of the other (potential energy barrier is not crossed). On the other hand, if one or two real positive roots $r_m$ do exist, one entity orbits the other with a minimum center-to-center distance $r_m$ corresponding to the larger of possible two positive roots (potential energy barrier is crossed). In the latter case, a collision will happen only if the minimum center-to-center distance is smaller than the sum of entities' hard sphere radii. The determination of the critical impact parameter $b_c$ requires a detailed discussion on which scenario is to happen (see Materials and Methods), which ultimately leads to:

$$b_c^2 = \begin{cases} \omega_v(R_m), & \text{if } R_m > R_i + R_j \\ \omega_v(R_i + R_j), & \text{if } R_m \leq R_i + R_j \end{cases}. \tag{6}$$

Eq. 6 is valid for attractive potentials in the functional form of $U(r_m) \propto -r_m^{-n}$. The results for repulsive potentials or potentials in other functional forms may be slightly different, but the analysis follows the same logic. A simple example for Eq. 6 is the Coulomb potential of oppositely charged particles: the manner it scales with separation distance, i.e., $U(r) \propto -r^{-1}$, ensures that the condition $R_m = 0 \leq R_i + R_j$ always holds regardless of the particle and charge properties, so the solution is further simplified to $b_c^2 = \omega_v(R_i + R_j)$. This special case of solution has been invoked commonly for modelling collision rate coefficients involving ions and charged particles (38, 39).

**Monomer-monomer collisions**. Vapors relevant to atmospheric clustering are mostly composed of polar and polarizable molecules interacting through Van der Waals force scaling with the separation distance based on $U(r) \propto -\alpha r^{-6}$, and the pre-factor $\alpha$ is expected to be larger for cases of permanent dipole interactions compared to cases involving induced or instantaneous dipole interactions (see our note (40) and the reference (41) for the exact expressions of $\alpha$ in different cases). Thus, we represent the monomer-monomer interaction using the general form of the attractive Van der Waals potential:

$$U_{mm}(r) = -4\epsilon \left(\frac{\sigma}{r}\right)^6, \qquad (7)$$

where $\epsilon$ represents the depth of the potential well, $\sigma$ is a characteristic length, and the subscript "$mm$" denotes the monomer-monomer interaction. Note that the repulsive part of Van der Waals potential describing Pauli repulsion, which requires monomer centers to be very close to each other, has minimal influence on trajectories prior to collision (e.g., for $r > R_i + R_j$), and is hence neglected. Values for $\epsilon$ and $\sigma$ are readily available from literature depending on the type of monomers. Here, we obtained the $\epsilon$ and $\sigma$ parameters by computing the potential of mean force (PMF) using a metadynamics simulation (42) with a validated all-atom force field (33, 43) (results and computational details are provided in the Supporting Information, Table S1 and Fig. S1). Substituting Eq. **7** into Eqs. **5** and **6** to replace $U(r_m)$ with $U_{mm}(r_m)$, we found that $\omega_v(r_m, U_{mm})$ has a positive minimum $\omega_v(R_m, U_{mm}) = 3\sqrt[3]{2}\sigma^2 \cdot [\epsilon/(m_{ij}v^2)]^{1/3}$, located at $R_m(U_{mm}) = \sqrt[3]{4}\sigma \cdot [\epsilon/(m_{ij}v^2)]^{1/6}$. With known set of monomer types, $b_c$ is readily evaluated from Eq. **6** with respect to any given relative speeds $v$.

We validated the prediction of Eqs. **5 - 7** with results from molecular dynamics simulations, using the collision between the sulfuric acid monomer $H_2SO_4$ and the acid-base monomer $[(CH_3)_2NH_2^+ \cdot HSO_4^-]_1$ as an example. The molecular dynamics collision simulations were performed using the all-atom force field parametrized by Loukonen et al. (33), where interaction parameters for sulfuric acid, bisulfate, and dimethylammonium were fitted based on the Optimized Potentials for Liquid Simulations all-atom procedure (43) (computational details are described in the Supporting Information). Briefly, we equilibrated a pair of monomers at a target temperature $T = 300$ K under a NVT ensemble. Once equilibration was reached, a set of $b$ and $v$ pair was assigned to the center of mass of the molecules, and their trajectories were then iterated under a NVE ensemble. A collision event was identified if their center of mass distance was within the sum of the hard sphere radii $(R_i + R_j)$ for at least one time frame during the simulation. This procedure for the same $b$ and $v$ pair was repeated until statistical convergence was achieved, and collision probability $P_c(b, v)$ was obtained based on the identified number of successful collision events, as plotted in Fig. 1A. To compare the molecular dynamics simulation results with the theoretical prediction, we retrieved the points in Fig. 1A fulfilling the condition $P_c(b_c, v) = 0.5$, and plot them in Fig. 1B against the predictions of Eqs. **5 - 7**. Also shown are data from "trajectory simulations" where trajectories of pointlike particles $H_2SO_4$ and $[(CH_3)_2NH_2^+ \cdot HSO_4^-]_1$ were numerically solved with the Velocity-Verlet algorithm where the potential was given by Eq. 7 (with the same PMF based $\epsilon$ and $\sigma$ values as used in the theoretical predictions), e.g., $m_i \frac{d^2 \vec{r}_i}{dt^2} = -\frac{dU_{mm}}{dr} \cdot \frac{\vec{r}_i - \vec{r}_j}{|\vec{r}_i - \vec{r}_j|}$. As expected, the theoretical approach predicts a set of critical impact parameters $b_c$ that agrees perfectly with those from trajectory simulations, and both agree well with $P_c(b_c, v) = 0.5$ data from molecular dynamics simulations.

**Monomer-cluster collisions**. Predicting monomer-cluster collision rates using Eqs. **3** and **5 – 7** requires an appropriate description of corresponding potentials, which may be computed for any monomer-cluster combinations from molecular dynamics or more accurate *ab initio* methods. However, without loss of generality, we obtain the monomer-cluster potentials $U_{mc}(r)$ through integrating the monomer-monomer potential over the cluster's volume, assuming uniform monomer number density:

$$U_{mc}(r) = \oiiint_{V_c} U_{mm}(r)\, \rho_c\, dV = -\frac{4 n_c \epsilon \sigma^6}{(r^2 - R_c^2)^3}, \tag{8}$$

where $\rho_c$ is the monomer number density, $R_c$ is the radius of the cluster, $V_c$ is the volume of the cluster, and $n_c = \rho_c V_c$ is the total number of monomers in the cluster. Eq. **8** is valid for clusters composed of the same type of monomers, and the results for clusters composed of a mixture of different monomers can be found in the Supporting Information. Note that further integrating $U_{mc}(r)$ over the volume of the other cluster recovers the well-known results of Hamaker (44) for Van der Waals attractive potential between two spherical objects, which can be invoked effectively to treat cluster-cluster collision rates. Using integrated cluster potential has been previously done, but to the best of our knowledge a generic, consistent, and easy to implement model has not yet been developed to the best of our knowledge; previous models usually resort to fittings against experimental data (45-47). Our current theoretical model requires only the elementary monomer-monomer interactions as a-priori information, and it is thus generic. Cluster-cluster collisions possess distinct features compared to monomer-cluster collisions, which may have significant implications on the behavior of systems involving cluster/nanoparticle coagulations. We will further discuss these features in future publications. Replacing $U(r_m)$ in Eq. **5** with $U_{mc}(r_m)$, $\omega_v(r_m, U_{mc})$ was once again found to be a concave function with a positive minimum (note that here we require $r_m > R_c$). By taking the derivative of $\omega_v(r_m, U_{mc})$ with respect to $r_m$, it was further found that the minimum is located at $r_m = R_m(U_{mc})$ and satisfies:

$$\sum_{i=0}^{4} a_i R_m^{2i} = 0, \tag{9}$$

with $a_0 = R_c^2(R_c^6 - l_c^6)$, $a_1 = -2l_c^6 - 4R_c^6$, $a_2 = 6R_c^4$, $a_3 = -4R_c^2$, and $a_4 = 1$, where $l_c \equiv [8n_c \epsilon \sigma^6/(m_{ij} v^2)]^{1/6}$ is a cluster size- and potential-dependent characteristic length. Eq. **9** is nothing but a fourth order (quartic) equation after writing it in terms of $R_m^2$, and it can be shown that only one real root exists for $R_m > R_c$ (see Supporting Information). The solution is readily available:

$$R_m^2(U_{mc}) = R_c^2 + M + \sqrt{-M^2 - \frac{q}{4M}}, \tag{10}$$

where $q = -2l_c^6$, $M = \frac{1}{2}\sqrt{\frac{1}{3}(N + \frac{\Delta_0}{N})}$, and $N = \sqrt[3]{\left(\Delta_1 + \sqrt{\Delta_1^2 - 4\Delta_0^3}\right)/2}$ with $\Delta_0 = -36R_c^2 l_c^6$ and $\Delta_1 = 108 l_c^{12}$. With Eqs. **6** and **10**, we can predict $b_c$ theoretically for any given set of monomer-cluster combinations at any given relative speed $v$, with $\epsilon$ and $\sigma$ taken from the corresponding monomer-monomer PMF from Table S1. As in the case of the monomer-monomer collisions, again this theoretical framework of predicting $b_c$ was validated by comparing the results of Eqs. **6** and **10** to trajectory simulations for monomer-cluster collisions (see Fig. S3 for details).

The resulting $b_c$ resulting from Eqs. **6** and **10** was substituted into Eq. **3** where the integral was numerically evaluated to obtain the monomer-cluster collision rates which are shown (as solid lines) in Fig. 2A and 2B for (A) $H_2SO_4$ & $[(CH_3)_2NH_2^+ \cdot HSO_4^-]_n$ collisions and (B) $(CH_3)_2NH$ &

$[(CH_3)_2NH_2^+ \cdot HSO_4^-]_n$ collisions at temperatures in an atmospherically relevant range ($T = 200$, 300, and 400 K). Also shown in Fig. 2A and 2B are collision rates from molecular dynamics simulations (red dots) calculated by integrating the collision probability $P_c(b,v)$ over all possible impact conditions ($b$ and $v$) sampled from corresponding Maxwell-Boltzmann distribution at target temperatures, i.e.:

$$k_{i,j}^{MD} = 2\pi \int_0^\infty dv \int_0^\infty db\, v\, f(v)\, b\, P_c(b,v), \tag{11}$$

where the superscript "MD" represents results from molecular dynamics. Note that Eq. **11** is reduced to Eq. **3** by assuming $P_c(b,v) = 1$ for $b < b_c$ and $P_c(b,v) = 0$ for $b > b_c$, which turns out to be reasonable as confirmed by the sharp transition of $P_c(b,v)$ from 1 to 0 at the region near the theoretical $b_c(v)$ line in Fig. 1A. The agreement of the molecular dynamics results with theoretical data is demonstrated in Fig. 2A and 2B at various system temperatures by showing the cluster size dependence of both the collision rate and enhancement factor $\eta$ defined as the ratio between the collision rate of interacting clusters and the hard-sphere counterparts.

**Collision-sticking kinetics**. We used the molecular dynamics data to retrieve the sticking rate, i.e., the rate of successful collision events leading to the formation of stable product clusters. The sticking probability is defined similarly as collision probability, based on the identified number of events where stable product clusters are formed:

$$s_{i,j}^{MD} = 2\pi \int_0^\infty dv \int_0^\infty db\, v\, f(v)\, b\, P_s(b,v). \tag{12}$$

To invoke Eq. **12**, a consistent criterion of physical significance needs to be proposed to parameterize the sticking probability. Note that there are two distinct types of monomer-cluster interactions that play important roles in the cluster formation process: (1) the vibrational coupling of the condensable vapor molecule onto the cluster surface right after a collision, which, if successful, drives the cluster formation while storing excess energy (48) due to bond formation, and (2) the collision between background inert carrier gas molecules and clusters, which dissipates the excess energy of nascent clusters and equilibrates/thermalizes the clusters to a temperature corresponding to that of the carrier gas. We considered the following scenario: following a condensable vapor-cluster collision, the vapor monomer experiences a few rounds of unsteady vibrations on the cluster surface before forming stable hydrogen bonds (sticking) or dissociating; in the former case, a stable cluster with excess thermal energy is produced. At this point, the only possible reason for further loss of the monomer is thermal fluctuation, i.e., evaporation, as the monomer has become thermally indistinguishable from other molecules in the cluster. At typical atmospheric conditions, the time scale $t_s$ for the formation of stable bonds after a condensable vapor-cluster collision is smaller than the mean free time $t_c$ of the background inert carrier gas-cluster collision and hence the time $t_{eq}$ required to fully thermalize the cluster; note that $t_{eq} \sim 10 t_c$ as the thermalization process only requires tens of collisions for the small clusters discussed here (49). Moreover, $t_s$, $t_c$, and $t_{eq}$ can be assumed to be orders of magnitude smaller than the time scale for a monomer to evaporate ($t_{ev}$) (see the schematic diagram in Fig. 2). For example, in the case of a $[H_2SO_4]_2$ cluster (hard sphere radius $R_c \approx 0.3$ nm) surrounded by air at the standard atmospheric pressure at 300 K: $t_s \sim 2\, L_c/\bar{c} \sim 10$ ps estimated based on its molecular neighbor separation distance $L_c$ and the mean thermal speed $\bar{c}$, $t_c \sim 10^2$ ps calculated based on the molecular collision frequency (hence $t_{eq} \sim 10^3$ ps), and $t_{ev} \sim 10^6$ ps estimated based on the free molecule regime evaporation rate as described in Supporting Information; note that, due to release of the latent heat, the nascent dimer may reach an "effective temperature" of about 350 K to 400 K corresponding to $t_{ev} \sim 10^4$ ps to $10^5$ ps which are still much larger than $t_s$, $t_c$, and $t_{eq}$.

These time scales strongly indicate that a stable cluster will form if the impinging condensable vapor monomer can survive a very short window of unsteady vibration right after a collision. Based on this, we monitored the location of the impinging monomer and marked a successful sticking event if it stayed on cluster surface for more than 25 or alternatively 50 picoseconds. The results of sticking probability (Fig. S2) and sticking rates (Fig. 2, circles and squares) calculated at these two threshold times match each other very well, suggesting that stable clusters are indeed formed. When comparing $H_2SO_4$ - $[(CH_3)_2NH_2^+ \cdot HSO_4^-]_n$ sticking rates with collision rates (Fig. 2A), no visible difference can be observed: a collision essentially always leads to a sticking. However, the situation is clearly different for the case of $(CH_3)_2NH$ - $[(CH_3)_2NH_2^+ \cdot HSO_4^-]_n$ collisions (Fig. 2B), attributed to the fact that the strength of hydrogen bonding between $(CH_3)_2NH$ and $[(CH_3)_2NH_2^+ \cdot HSO_4^-]_n$ is weaker than that between $H_2SO_4$ and $[(CH_3)_2NH_2^+ \cdot HSO_4^-]_n$. This deviation between the collision and sticking rates is enhanced with increasing temperature which effectively enhances the magnitude of thermal vibrations. At a system temperature less than 300 K, the deviation is expected to be solely a consequence of the vibrational coupling process described by "1." and "2." in the schematic in Fig. 2C based on the corresponding time scales. However, at 400 K visible differences between sticking rates recorded at 25 and 50 picoseconds appear, indicating that the temperature is approaching the point where $t_{ev}$ of the nascent cluster has decreased to a value comparable to $t_s$. At this point, the deviation between collision and sticking rates defined using these threshold values is caused by the coupled effects of the vibrational coupling process and the evaporation of the nascent cluster. Moreover, the difference between sticking and collision rates decreases with increasing cluster size: larger sizes favor formation of stable clusters. This is not surprising because larger clusters offer more potential sites for hydrogen bonding and molecular neighbors to enhance mutual attractions. They also serve as more efficient sinks for dissipating the latent heat released during bond formation upon collisions (49, 50), directly enhancing the probability of sticking. Fig. S4 of the Supporting Information shows the results for the collision-sticking kinetics between $H_2SO_4$ and $[H_2SO_4]_n$ clusters; the $H_2SO_4$ monomer has much higher sticking probability on $[(CH_3)_2NH_2^+ \cdot HSO_4^-]_n$ clusters than on pure $[H_2SO_4]_n$ clusters at small *n* values, providing direct molecular evidence on how base species enhances the formation of the smallest clusters(51). The observed collision-sticking kinetics and the criterion proposed here can serve as solid foundations for developing analytical sticking rate models based on comparing relative magnitudes of monomer-cluster potential interaction and the monomer impinging kinetic energy.

As mentioned previously, the proton transfer between acid and base molecules modifies dipole moments of products, which enhances both the products' stability and their capability of capturing free monomers. To get a sense to what extent do these two effects contribute to the rate of cluster formation, we devised a set of molecular dynamics collision simulations using clusters where proton transfer was prohibited, i.e., clusters were forced to be of the form $[(CH_3)_2NH \cdot H_2SO_4]_n$. The practical differences between clusters with and without proton transfers are: (1) in the molecular dynamics simulation, the atoms in $[(CH_3)_2NH \cdot H_2SO_4]_n$ and $[(CH_3)_2NH_2^+ \cdot HSO_4^-]_n$ clusters were assigned different partial charges to mimic corresponding dipole moments, and (2) in the theoretical prediction, $[(CH_3)_2NH_2^+ \cdot HSO_4^-]_n$ clusters were treated as the assemble of $(CH_3)_2NH_2^+ \cdot HSO_4^-$ monomers while $[(CH_3)_2NH \cdot H_2SO_4]_n$ clusters were treated as the mixture of $(CH_3)_2NH$ and $H_2SO_4$ monomers with uniform spatial distributions. Proton transfer may happen fully or partially in the cluster depending on the relative abundance of free acid and base molecules. Nonetheless, current simulations correspond to the two limiting cases with proton transfer being fully realized or completely prohibited. Theoretical $H_2SO_4$- and $(CH_3)_2NH$ - $[(CH_3)_2NH \cdot H_2SO_4]_n$ collision rates at 300 K obtained from treating $[(CH_3)_2NH \cdot H_2SO_4]_n$ as mixture of $(CH_3)_2NH$ and $H_2SO_4$ monomers, and sticking rates retrieved from molecular dynamics data following the same procedure as those in fully proton-transferred cases are provided in the Supporting Information (Fig. S5, details of the theoretical procedure are also provided in the same section). Compared with fully proton-transferred cases, the collision rates in non-proton-transferred cases are only slightly decreased but the sticking rates are notably smaller, especially for cases of $(CH_3)_2NH$ - $[(CH_3)_2NH \cdot H_2SO_4]_n$ collisions. Overall, it can be concluded that the increased strength of potential caused by proton transfer increases cluster formation rates mainly through increasing the sticking probability and cluster stability, while the contribution from the increased collision rate plays a minor role.

Fig. 3 shows the theoretical predictions of the enhancement factor $\eta$ against hard sphere collision rate coefficients as a function of cluster radius for all four cases at 300 K, starting from the smallest cluster composed of only one molecule. It was found that the monomer type has minimal influence on the collision enhancement factor: the effects of stronger interactions for cases involving $H_2SO_4$ are cancelled out by larger mean thermal speed for cases involving $(CH_3)_2NH$. All clusters gradually approach to the hard sphere behavior as their size increases, i.e., $\eta \to 1$ as $n \to \infty$. Our novel theoretical approach for calculating molecular collision rates is generic and can be invoked directly in the kinetic equation Eq. **1** to predict atmospheric clustering of non-volatile compounds proceeding at the kinetic limit (34, 51, 52). Clustering involving volatile compounds are expected be sensitive also to rates of evaporation and mass accommodation (or sticking probability), and the modelling requires extra efforts to include such effects but can be assisted by the collision-sticking kinetics unraveled in this study.

**Materials and Methods**

**Properties of the function $\omega_v(r_m)$.** The solution of the critical impact parameter $b_c$ from Eq. **6** largely depends on properties of the function $\omega_v(r_m)$ that in turn is affected by the form of intermolecular potentials. Let us consider a representative type of potential $U_n(r) \propto -\alpha r^{-n}$ covering a common group of neutral or charged molecular interactions, including integrations of ion-ion ($n = 1$), ion-dipole ($n = 4$), and Van der Waals ($n = 6$, for dipole-dipole, dipole-induced dipole, and dispersion). Substituting this form for the potential into $\omega_v(r_m)$, we obtained:

$$\omega_v(r_m, U_n) = r_m^2 \left(1 + \frac{2\alpha}{m_{ij}v^2} \cdot \frac{1}{r_m^n}\right), \tag{13}$$

where the pre-factor $\alpha$ is depended on the type of potentials and is positive for all attractive interactions. For $n = 1$ or 2, it is straightforward to find that the criterion $R_m = 0 \leq R_i + R_j$ is fulfilled, such that the solution of $b_c$ from Eq. **6** is simplified to: $b_c^2 = \omega_v(R_i + R_j, U_n)$. For $n \geq 3$, by applying the inequality of arithmetic and geometric means to the right-hand side of Eq. **13**, we found that:

$$\omega_v(r_m, U_n) = \frac{r_m^2}{n-2} + \cdots + \frac{r_m^2}{n-2} + \frac{\alpha}{m_{ij}v^2} \cdot \frac{1}{r_m^{n-2}} + \frac{\alpha}{m_{ij}v^2} \cdot \frac{1}{r_m^{n-2}} \geq n \sqrt[n]{\frac{1}{(n-2)^{n-2}} \cdot \left(\frac{\alpha}{m_{ij}v^2}\right)^2} =$$

$$\frac{n}{n-2} \left[\frac{\alpha(n-2)}{m_{ij}v^2}\right]^{\frac{2}{n}}. \tag{14}$$

We therefore obtained the minimum $\omega_v(R_m, U_n) = \frac{n}{n-2}\left[\frac{\alpha(n-2)}{m_{ij}v^2}\right]^{2/n}$ located at $R_m(U_n) = \left[\frac{\alpha(n-2)}{m_{ij}v^2}\right]^{1/n}$, and hence Eq. **6** is expressed more explicitly as: $b_c^2 = \omega_v(R_i + R_j, U_n)$ if $R_i + R_j \geq R_m(U_n)$, and $b_c^2 = \omega_v(R_m, U_n)$ if $R_i + R_j < R_m(U_n)$. Fig. **4A** shows the results of $b_c$ for collisions of (left panel) $(CH_3)_2NH_2^+$ & $HSO_4^-$ ions: attractive Coulomb potential with $n = 1$ and $\alpha = q_i q_j/(4\pi\varepsilon_0)$ assuming vacuum permittivity, and (right panel) $H_2SO_4$ & $(CH_3)_2NH_2^+ \cdot HSO_4^-$ neutral molecules: Van der Waals potential with $n = 6$ and $\alpha = 4\epsilon\sigma^6$ taken from Table S1.

**Determination of the critical impact parameter $b_c$.** For the attractive Coulomb potential, function $\omega_v(r_m, U_n)$ increases monotonically from 0 to $+\infty$ in the interval of $r_m \geq 0$ and one real positive root $r_m$ always exists for Eq. **5**, therefore the energy barrier can always be crossed regardless of impact conditions and the situation is straightforward. The case of the Van der Waals potential, however, is complicated by the feature of function $\omega_v(r_m, U_n)$. There are two possible scenarios depending on the relative value of $R_m$ and $R_i + R_j$ corresponding to the schematics in Fig. 4B and 4C (applicable generally to all $n \geq 3$ cases), where the blue circular area represents the area of influence with a radius of $R_i + R_j$, the solid black line represents the relative trajectory before the collision, and the dotted part is the imaginary trajectory if the colliding partners are treated as point particles: (1) when the relative speed $v$ is small, $R_m > R_i + R_j$ fulfills (Fig. 4B). There is a potential energy barrier initially, which can be crossed by increasing the impact parameter $b$. At the point the barrier is crossed, we have $r_m = R_m > R_i + R_j$, indicating that collisions will no longer happen even if the impact parameter $b$ is further increased. Hence, the critical impact parameter is determined by $b_c^2 = \omega_v(R_m)$. (2) Once the relative speed $v$ is large enough to lead to $R_m < R_i + R_j$ (Fig. 4C), then at the point where the potential energy barrier is crossed, we have $r_m = R_m < R_i + R_j$, suggesting that there is still room for collisions by further increasing the impact parameter $b$ until $r_m = R_i + R_j$. Hence, the critical impact parameter is determined by $b_c^2 = \omega_v(R_i + R_j)$. Note that in both cases we have dropped the smaller positive root $r_m$ of Eq. **5**, resulted from the decreasing (dotted) part of function $\omega_v(r_m, U_n)$; the choice is made because $r_m$ should increase rather than decrease with increasing the impact parameter $b$, and $r_m$ should approach infinite rather than zero as the impact parameter $b$ approaches infinite. The feature of function $\omega_v(r_m, U_n)$ for the Van der Waals potential leads to the situation that the solution of $b_c^2$ transits from $\omega_v(R_m)$ to $\omega_v(R_i + R_j)$ as the relative speed increases, which is validated by the excellent agreement between theoretical predictions and trajectory simulations shown in Fig. 4A. Note that the transition between modes for the $H_2SO_4$ & $(CH_3)_2NH_2^+ \cdot HSO_4^-$ collision happens at a relative speed of ~800 $m\ s^{-1}$, above which accounts for less than 1% of the relative speeds for the investigated cases according to the Maxwell-Boltzmann speed distributions. Therefore, the $\omega_v(R_m)$ collision mode can already approximate the collision rates well. However, this may not be applicable universally; for collision of weakly bonded molecules, e.g., water, smaller values of $\epsilon$ can cause the transition to happen at a low relative impact speed which overlaps heavily with corresponding velocity distributions, such that both modes need to be carefully considered.

**Supporting Information**. Details of the PMF calculation, molecular dynamics collision simulation setup, derivation of the root of Eq. **9**, Estimation of the time scale for monomer evaporation and gas-cluster collision, validation of the theoretical prediction of critical impact parameter from Eqs. **6** and **10**, collision and sticking rates for $H_2SO_4$ & $[H_2SO_4]_n$ pairs, and collision and sticking rates for $H_2SO_4$ & $[(CH_3)_2NH \cdot H_2SO_4]_n$ or $(CH_3)_2NH$ & $[(CH_3)_2NH \cdot H_2SO_4]_n$ pairs are provided in the Supporting Information.

**Acknowledgments**

This work was completed with the financial support of the ERC grant no. 692891-DAMOCLES, the University of Helsinki, Faculty of Science ATMATH project, and the Academy of Finland flagship program (grant nr. 337549). We thank the CSC – Finnish IT Centre for access to computer clusters and also computer capacity from the FinnishGrid and Cloud Infrastructure (persistent identifier urn:nbn:fi:research-infras-2016072533).

**Figures**

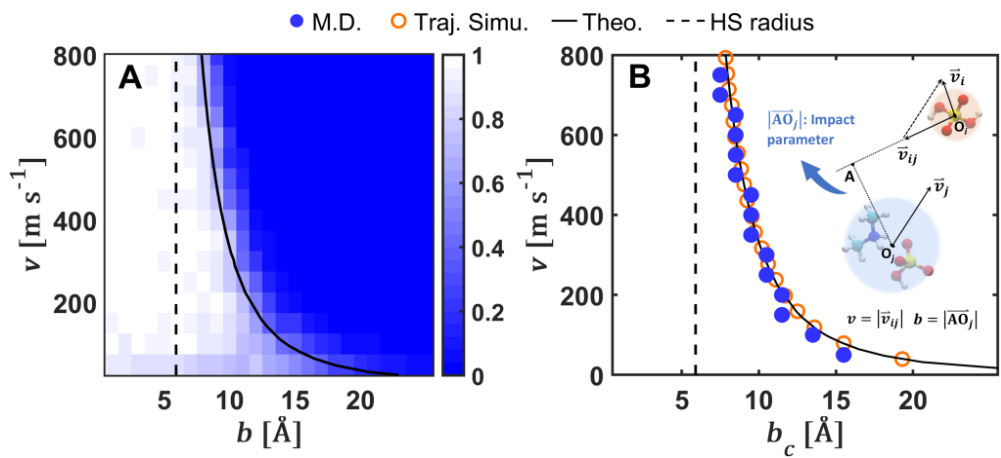

**Figure 1**. (A) Collision probability $P_c$ and (B) critical impact parameter $b_c$ for the collision pair of $H_2SO_4$ and $[(CH_3)_2NH_2^+ \cdot HSO_4^-]_1$. The molecular dynamics (M.D.) data of $b_c$ in B are retrieved from A where $P_c(b_c, v) = 0.5$ is satisfied. Trajectory simulation (Traj. Simu.) results of $b_c$ are based on numerical solutions of corresponding point particle trajectories driven by the Van der Waals attractive force. The solid line represents theoretical predictions (Theo.) and the dotted line represents the sum of the hard sphere radii of $H_2SO_4$ and $[(CH_3)_2NH_2^+ \cdot HSO_4^-]_1$.

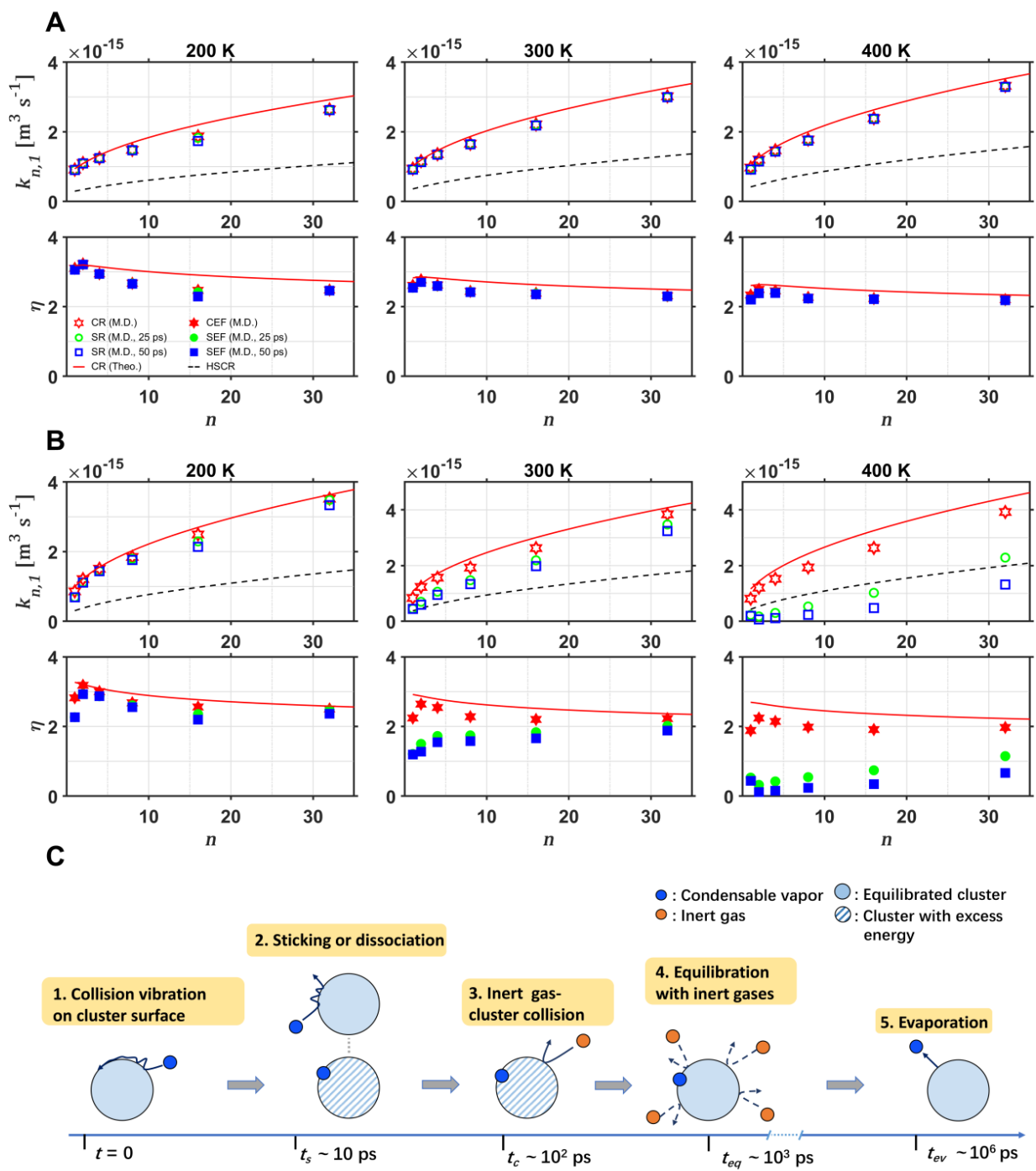

**Figure 2**. Collision rates (CR), sticking rates (SR), hard sphere collision rates (HSCR), collision rate enhancement factor (CEF = CR/HSCR), and sticking rate enhancement factor (SEF = SR/HSCR) for collision pairs of (A) $H_2SO_4$ & $[(CH_3)_2NH_2^+ \cdot HSO_4^-]_n$ and (B) $(CH_3)_2NH$ & $[(CH_3)_2NH_2^+ \cdot HSO_4^-]_n$ at 200, 300, and 400 K from theoretical predictions (Theo.) and molecular dynamics simulations (M.D. at 25 and 50 ps). (C) The schematic shows the time scale for events of significance.

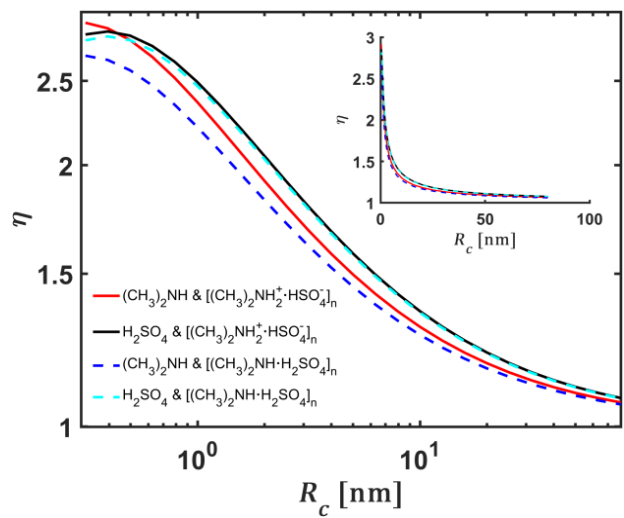

**Figure 3.** Collision enhancement factor for the four investigated cases. Enhancement factors gradually decay to 1 with the increase of cluster radius $R_c$, indicating a transition to hard sphere like behavior. The main plot shows the data with logarithmic radius axis, and the inset with linear radius axis.

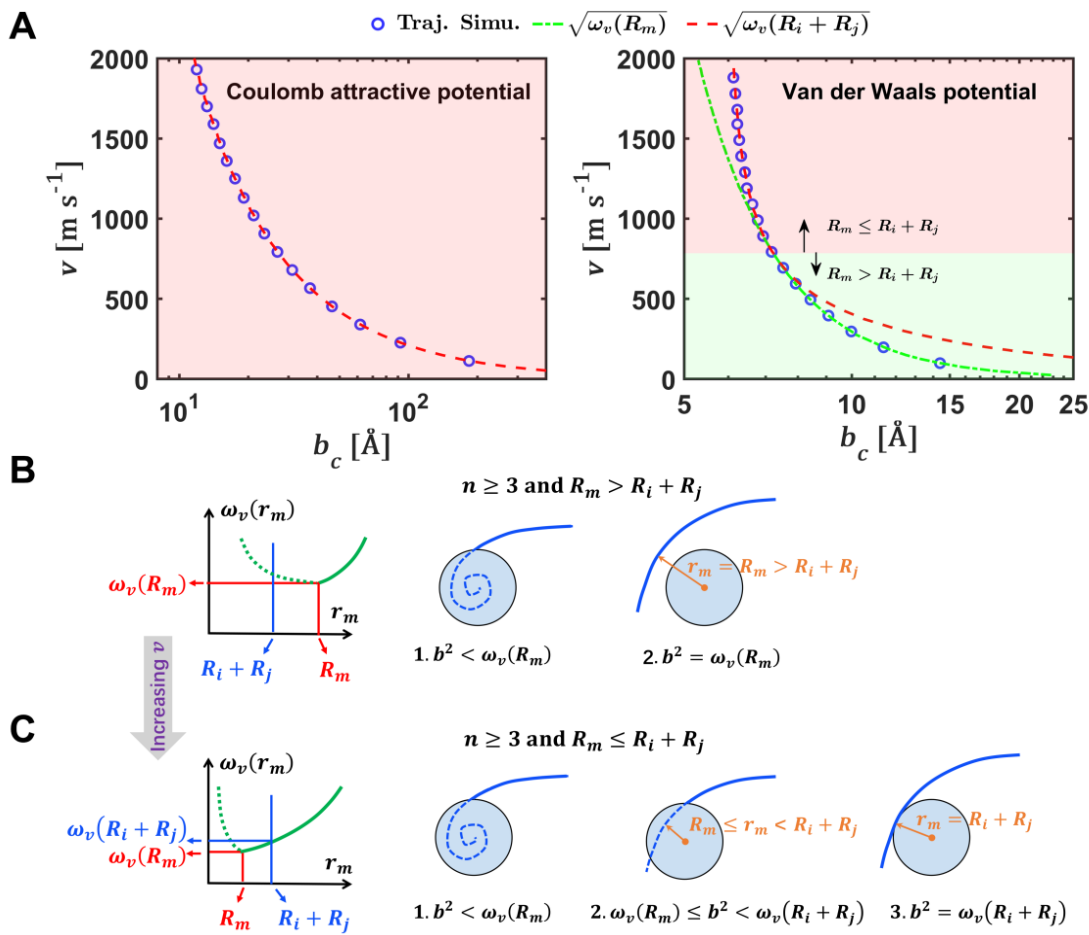

**Figure 4**. (A) Critical impact parameters for examples of Coulomb attractive potential $(CH_3)_2NH_2^+$ & $HSO_4^-$ pair and Van der Waals potential $H_2SO_4$ & $(CH_3)_2NH_2^+ \cdot HSO_4^-$ pair. The schematics show the two possible collision scenarios for the case of the Van der Waals potential: (B) $R_m > R_i + R_j$ and (C) $R_m < R_i + R_j$.

# Collision-sticking kinetics of acid-base clusters and its influence on atmospheric new particle formation

Huan Yang*[1], Bernhard Reischl[1], Ivo Neefjes[1], Valtteri Tikkanen[1], Jakub Kubečka[1], Theo Kurtén[2], and Hanna Vehkamäki[1]

[1]Institute for Atmospheric and Earth System Research/Physics, University of Helsinki, FI-00014 Helsinki, Finland;
[2]Institute for Atmospheric and Earth System Research/Chemistry, University of Helsinki, FI-00014 Helsinki, Finland;

*Corresponding Author: Huan Yang, Email: **huan.yang@helsinki.fi**

# Supporting Information

**Contents:**

- Molecular dynamics simulations (potential of mean force and collision simulation)

- The root of Eq. **9**

- Estimation of the time scale for monomer evaporation and gas-cluster collision

- Validation of Eqs. **6** and **10** against trajectory simulation and molecular dynamics

- Collision and sticking rates for $H_2SO_4$ & $[H_2SO_4]_n$ pairs

- Collision and sticking rates for $H_2SO_4$ & $[(CH_3)_2NH \cdot H_2SO_4]_n$ or $(CH_3)_2NH$ & $[(CH_3)_2NH \cdot H_2SO_4]_n$ pairs

**Molecular dynamics simulations**

**Potential of mean force (PMF) calculation.** To determine the thermally averaged interaction potential between two entities $i$ and $j$, we calculated the potential of mean force as a function of their center of mass distance from a well-tempered meta-dynamics simulation[1], using the PLUMED plug-in[2] for LAMMPS[3]. The OPLS all-atom[4] based force field parameters by Loukonen et al.[5] were used to describe the colliding molecules and clusters. We used the velocity Verlet integrator with a time step of 1 fs, where the Lennard-Jones interactions were cut off at 14 Å while electrostatic interactions were evaluated with a cut-off at 40 Å. 40 random walkers were employed. Gaussians with a width of 0.1 Å and initial height of $2k_BT$ [$k_BT/10$ for the $(CH_3)_2NH$ & $(CH_3)_2NH$ pair] were deposited every 500 steps along the collective variable; a harmonic wall was used to restrict the collective variable to values below 35 Å. A bias factor of 20 [5 for the $(CH_3)_2NH$ & $(CH_3)_2NH$ pair] was chosen, and a stochastic velocity rescaling thermostat (CSVR) with a time constant of 0.1 ps was used to maintain a constant temperature. Calculations were performed for five combinations of monomer pairs at 300 K, including: $(CH_3)_2NH$ & $(CH_3)_2NH$, $H_2SO_4$ & $H_2SO_4$, $H_2SO_4$ & $(CH_3)_2NH$, $H_2SO_4$ & $[(CH_3)_2NH_2^+ \cdot HSO_4^-]_n$, and $(CH_3)_2NH$ & $[(CH_3)_2NH_2^+ \cdot HSO_4^-]_n$. Note that we used the PMF at 300 K to approximate cases at 200 and 400 K in deriving corresponding collision rates; though there are slight differences in the PMF at these temperatures, their influence on the theoretical collision rates was found to be negligible. Results are summarized in Fig. S1 and Table S1, except that the $(CH_3)_2NH$ & $(CH_3)_2NH$ interaction was too small to be distinguished from thermal noise in our calculations, it was therefore directly taken as zero and is not shown in Fig. S1. The monomer-monomer interactions were assumed to obey the form of the Lennard-Jones potential such that the depth of the potential well (or binding energy) $\epsilon$ used in this work was directly taken as the minimum energy of the PMF curve and $\sigma$ used in this work was determined from $\sigma = r_\epsilon / \sqrt[6]{2}$, where $r_\epsilon$ is the corresponding distance for the minimum energy point on the PMF curve. The binding energy for sulfuric acid-sulfuric acid molecule is $-0.29$ eV which is in excellent agreement with previous *ab initio* results[6] of $-0.3$ eV obtained from Boltzmann averaging over four energy minimum structure dimers at 298.15 K, and in good agreement with recent calculations[7-8] of $-0.23$ and $-0.26$ eV at a higher level of theory.

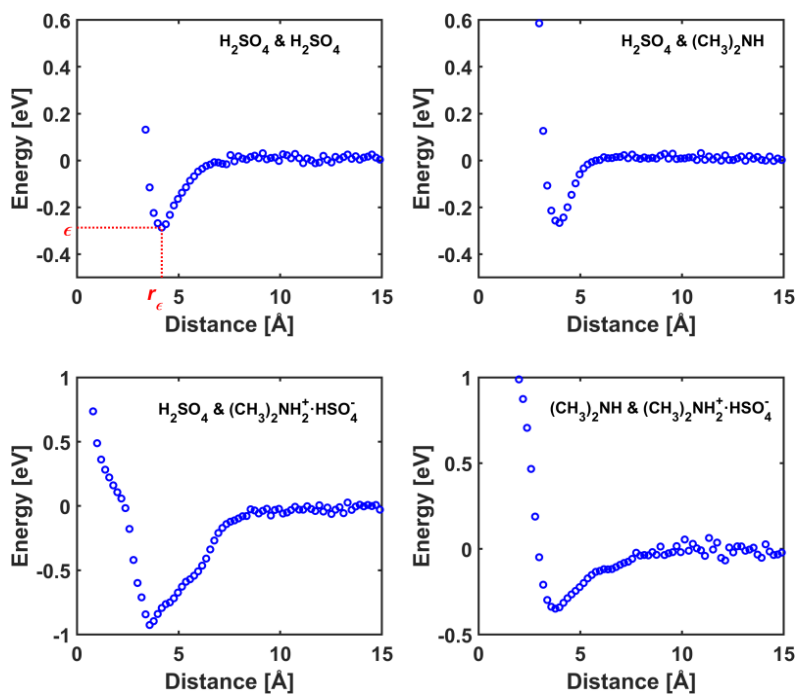

**Figure S1**. PMF for investigated monomer pairs.

Table S1. Calculated PMF parameters

|  | $(CH_3)_2NH$ & $(CH_3)_2NH$ | $H_2SO_4$ & $H_2SO_4$ | $H_2SO_4$ & $(CH_3)_2NH$ | $H_2SO_4$ & $(CH_3)_2NH_2^+ \cdot HSO_4^-$ | $(CH_3)_2NH$ & $(CH_3)_2NH_2^+ \cdot HSO_4^-$ |
|---|---|---|---|---|---|
| $\epsilon$ [eV] | ~0 | 0.29 | 0.26 | 0.93 | 0.35 |
| $\sigma$ [Å] | N/A | 3.71 | 3.54 | 3.19 | 3.36 |
| $r_\epsilon$ [Å] | N/A | 4.17 | 3.97 | 3.58 | 3.77 |

**Collision simulation setup.** Collision simulations were performed using LAMMPS with the force field parametrized by Loukonen et al.[5], where interaction parameters for sulfuric acid, bisulfate, and dimethylammonium were fitted based on the OPLS all-atom procedure. We have validated this set of force field parameters for molecular collision simulations in our previous work[9] by comparing predicted molecular structures, binding energies, and vibrational spectra of sulfuric acid molecules with *ab initio* results. Equations of motion were integrated using the Velocity Verlet algorithm with a time step of 1 fs, where Lennard-Jones interactions were cut off at 14 Å and electrostatic interactions were evaluated with a cut-off at 120 Å. In collision simulations, clusters composed of $n$ monomers were obtained by sintering and equilibrating two smaller clusters with $n/2$ monomers. The simulation setup is described as follows: initially, two colliding entities were placed at positions "infinitely far" from each other such that the mutual interaction was minimal, e.g. $\vec{r}_i = (0, 0, 0)$ and $r_j = (150, b, 0)$ Å, and their atomic velocities were sampled randomly from the Maxwell–Boltzmann distribution at a target temperature with the center-of-mass motion of each entity removed separately. Then the system was evolved for 100 ps, with a Nosé–Hoover[10-11] thermostat applied in the first 50 ps and removed for the remaining time; this procedure randomizes the intermolecular orientations and ensures equipartition of energy along the intramolecular degrees of freedom. At $t = 100$ ps, two center-of-mass translational velocities of equal magnitude but opposite directions were assigned to the two entities, i.e., $\vec{v}_i = (v/2, 0, 0)$ and $\vec{v}_j = (-v/2, 0, 0)$, to set them on a potential collision trajectory. Trajectories were then iterated for another ~200 ps until the formation of stable product or bounce off after collisions. Overall, $b$ and $v$ were selected uniformly from 50 to 800 m s$^{-1}$ in steps of 50 m s$^{-1}$ and 0 to 36 Å in steps of 1 Å, hence a total of more than 99 % of the Maxwell-Boltzmann distribution of the relative speed at the simulated temperatures lies within this range. For the same set of $b$ and $v$, $N$ (20 to 100 depending on cluster size) collision simulations were performed to balance the well-behaved statistics and computational cost. The influence of $N$ on the results is shown in Table S1. The resulting trajectories were then analyzed to determine the probability of collision and sticking; if colliding entities' center of mass distance was smaller than the summation of their hard-sphere radii for one time frame (or a given amount of time $t_s$) then a collision (or sticking) event was identified. This naturally leads to the probability of collision and sticking: $P_c(b, v) = N_c/N$ and $P_s(b, v) = N_s/N$, where $N_c$ and $N_s$ are identified numbers of collision and sticking events respectively. The collision probability and sticking probability retrieved for $t_s = 25$ and 50 ps for $H_2SO_4$ (upper panel) and $(CH_3)_2NH$ (lower panel) against $(CH_3)_2NH_2^+ \cdot HSO_4^-$ are shown below. As has been discussed in the main text, at least the following probabilities are almost indistinguishable from each other: 1) the collision probability and the sticking probability (both at 25 and 50 ps) for the case of the $H_2SO_4$ monomer, and 2) the sticking probability at 25 ps and the sticking probability at 50 ps for the case of the $(CH_3)_2NH$ monomer.

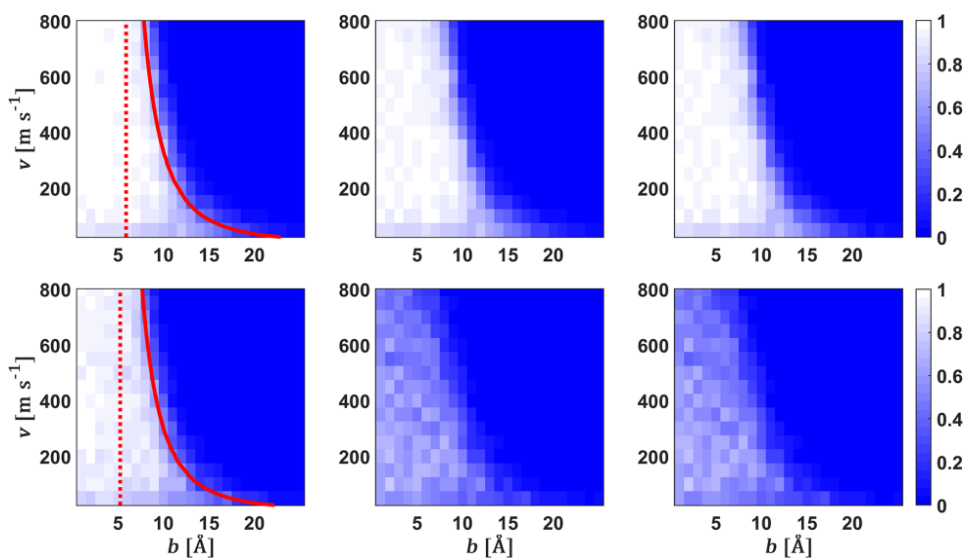

**Figure S2**. Collision $P_c$ and sticking probability $P_s$ for $H_2SO_4$ & $(CH_3)_2NH_2^+ \cdot HSO_4^-$ pair (upper panel) and $(CH_3)_2NH$ & $(CH_3)_2NH_2^+ \cdot HSO_4^-$ pair (lower panel). Left: collision probability, middle: sticking probability at 25 ps, and right: sticking probability at 50 ps. Red solid lines are $P_c(b_c, v) = 0.5$ and red dotted lines represents the sum of collision pairs' hard sphere radii.

**Table S2**. Dependence of results on the number of collision simulations performed at the same set of $b$ and $v$

| $N$ | Collision rate | Sticking rate (25 ps) | Sticking rate (50 ps) |
|---|---|---|---|
| 20 | 2.49 | 2.48 | 2.48 |
| 50 | 2.58 | 2.57 | 2.57 |
| 100 | 2.61 | 2.54 | 2.54 |
| 200 | 2.57 | 2.56 | 2.56 |
| 500 | 2.56 | 2.55 | 2.55 |
| 1000 | 2.56 | 2.55 | 2.55 |

**The root of Eq. 9**

In this part, we provide a reasoning to the statement that Eq. **9** always has exact one real root that is larger than $R_c$, and we shall identify that root. Substitution of Eq. **8** into $\omega_v(r_m)$ yields:

$$\omega_v(r_m, U_{mc}) = r_m^2 \left[1 + \frac{8n_c\epsilon\sigma^6}{m_{ij}v^2} \cdot \frac{1}{(r_m^2 - R_c^2)^3}\right]. \tag{S1}$$

We take the derivative of Eq. **S1** with respect to $r_m$ and obtain:

$$\omega_v'(r_m, U_{mc}) = 2r_m \left[1 - \frac{8n_c\epsilon\sigma^6}{m_{ij}v^2} \cdot \frac{2r_m^2 + R_c^2}{(r_m^2 - R_c^2)^4}\right] \equiv 2r_m[1 - \theta f(r_m)], \tag{S2}$$

where, $\theta = \frac{8n_c\epsilon\sigma^6}{m_{ij}v^2}$ is a positive constant and

$$f(r_m) = \frac{2r_m^2 + R_c^2}{(r_m^2 - R_c^2)^4} = \frac{2}{(r_m^2 - R_c^2)^3} + \frac{3R_c^2}{(r_m^2 - R_c^2)^4} \tag{S3}$$

decreases monotonically from $+\infty$ to 0 as $r_m$ increases from $R_c$ to $+\infty$. Therefore, $\omega_v'(r_m, U_{mc}) = 0 \Leftrightarrow f(r_m) = \frac{1}{\theta}$ should have a single root for $r_m > R_c$. Note that $\omega_v'(r_m, U_{mc}) = 0$ is equivalent to Eq. 9 which is a quartic function, so now it is safe to say that the maximum real root for this quartic function is $R_m(U_{mc})$, i.e.:

$$R_m^2 = -\frac{a_3}{4a_4} + M + \frac{1}{2}\sqrt{-4M^2 - 2p + \frac{q}{M}}, \tag{S4}$$

where $p = \frac{8a_4 a_2 - 3a_3^2}{8a_4^2} = 0$, $q = \frac{a_3^3 - 4a_4 a_3 a_2 + 8a_4^2 a_1}{8a_4^3} = -2l_c^6$, $M = \frac{1}{2}\sqrt{-\frac{2}{3}p + \frac{1}{3a_4}(N + \frac{\Delta_0}{N})}$, $N = \sqrt[3]{\frac{\Delta_1 + \sqrt{\Delta_1^2 - 4\Delta_0^3}}{2}}$, $\Delta_0 = a_2^2 - 3a_3 a_1 + 12a_4 a_0 = -36R_c^2 l_c^6$, and $\Delta_1 = 2a_2^3 - 9a_3 a_2 a_1 + 27a_3^2 a_0 + 27a_1^2 a_4 - 72a_4 a_2 a_0 = 108 l_c^{12}$ with $l_c \equiv \left(\frac{8n_c \epsilon \sigma^6}{m_{ij} v^2}\right)^{\frac{1}{6}}$. Substituting the expressions of coefficients $a_i$ into Eq. S4 and rearranging leads to Eq. 10 in the main text.

**Estimation of the time scale for monomer evaporation and gas-cluster collision**

The mean time between two subsequent monomer evaporation events was estimated based on the free molecular regime evaporation theory. The number flux ($J_{c,m}$) of monomers from the cluster surface is expressed as:

$$J_{c,m} = \pi R_c^2 c_m (n_e - n_\infty), \tag{S5a}$$

where $R_c$ is the radius of the cluster, $c_m = \sqrt{8k_B T/(\pi m_m)}$ is the mean thermal speed of the vapor monomer with $m_m$ being its molecular mass, $n_\infty$ is the background monomer number concentration, and $n_e$ is the equilibrium monomer number concentration at the nanoparticle surface, evaluated based on the Kelvin effect:

$$n_e = n_{ef} \exp\left(\frac{2\gamma v_m}{R_c k_B T}\right), \tag{S5b}$$

where $n_{ef} = P_{ef}/(k_B T)$ is the equilibrium monomer number concentration for flat surfaces with $P_{ef}$ being the vapor pressure for corresponding species, $\gamma$ is the nanoparticle surface tension, and $v_m$ is the monomer molecular volume. Therefore, the mean time between two subsequent monomer evaporation events for the case of the $H_2SO_4$ monomer from a $[H_2SO_4]_5$ cluster ($R_c \approx$ 0.5 nm) at 300 K under the vacuum condition ($n_\infty = 0$) was estimated as: $t_{ev} \sim 1/J_{c,m} \sim 10^6$ ps, with $\gamma = 0.0496\ N\ m^{-1}$ and $P_{ef} = 0.0023\ Pa$ taken from corresponding references[12-13].

The mean time between two subsequent background gas-cluster collisions was estimated based on their collision rate:

$$\beta_{c,g} = \pi R_c^2 c_g n_g, \tag{S6}$$

where $\beta_{c,g}$ is the gas-cluster collision rate, $c_g = \sqrt{8 k_B T/(\pi m_g)}$ is its mean thermal speed of the background gas with $m_g$ being its molecular mass, and $n_g = P_g/(k_B T)$ is its concentration, but note that the motion of cluster and the radius of the gas molecule has been neglected for convenience purpose. Therefore, for the case of, e.g., the $[H_2SO_4]_5$ cluster ($R_c \approx 0.5\ nm$) surrounded by $N_2$ gases at 300 K at the standard atmospheric pressure ($P_g = 101325$ Pa), the mean time for two subsequent collisions was estimated as: $t_c \sim 1/\beta_{c,g} \sim 10^2$ ps.

## Validation of Eqs. 6 and 10 against trajectory simulation and molecular dynamics

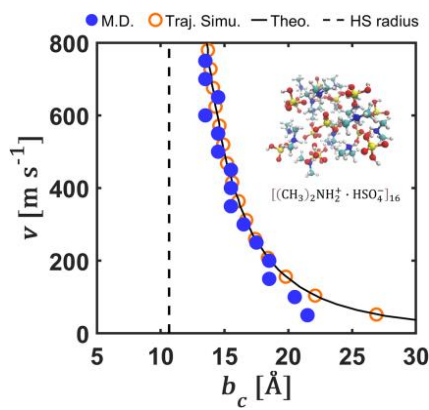

**Figure S3**. Critical impact parameter $b_c$ from molecular dynamics simulation (M.D.), trajectory simulation (Traj. Simu.), and theoretical prediction (Theo.) from Eqs. 6 and 10 for the $H_2SO_4$ and $[(CH_3)_2NH_2^+ \cdot HSO_4^-]_{16}$ collision pair. The molecular dynamics data of $b_c$ are retrieved at points in collision probability where $P_c(b_c, v) = 0.5$. The dotted line represents the hard sphere sum of the $H_2SO_4$ radius and $[(CH_3)_2NH_2^+ \cdot HSO_4^-]_{16}$ radius (HS radius).

The trajectory simulation results were obtained by numerically solving the equation of motion: $m_i \frac{d^2 \vec{r}_i}{dt^2} = -\frac{dU_{mc}}{dr} \cdot \frac{\vec{r}_i - \vec{r}_j}{|\vec{r}_i - \vec{r}_j|}$, with $U_{mc}(r) = -\frac{4n_c \epsilon \sigma^6}{(r^2 - R_c^2)^3}$ where $n_c$ and $R_c$ are based on the $[(CH_3)_2NH_2^+ \cdot HSO_4^-]_{16}$ cluster.

## Collision and sticking rates for $H_2SO_4$ & $[H_2SO_4]_n$ pairs

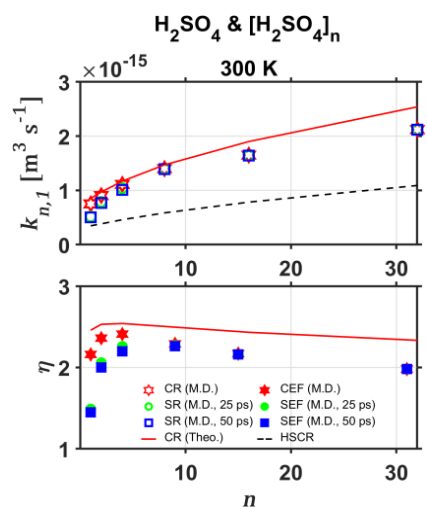

**Figure S4**. Collision rates (CR), sticking rates (SR), hard sphere collision rates (HSCR), collision rate enhancement factor (CEF = CR/HSCR), and sticking rate enhancement factor (SEF = SR/HSCR) for $H_2SO_4$ & $[H_2SO_4]_n$ pairs at 300 K from theoretical predictions (Theo.) and molecular dynamics simulations (M.D. at 25 and 50 ps).

**Collision and sticking rates for the $H_2SO_4$ & $[(CH_3)_2NH \cdot H_2SO_4]_n$ or $(CH_3)_2NH$ & $[(CH_3)_2NH \cdot H_2SO_4]_n$ pair**

The theoretical solution for monomer-cluster collision rates between a $(CH_3)_2NH$ or $H_2SO_4$ monomer and $[(CH_3)_2NH \cdot H_2SO_4]_n$ clusters were derived by assuming that, in $[(CH_3)_2NH \cdot H_2SO_4]_n$ clusters, $(CH_3)_2NH$ and $H_2SO_4$ molecules are uniformly distributed. That way, we integrated $(CH_3)_2NH$ - $(CH_3)_2NH$ & $(CH_3)_2NH$ - $H_2SO_4$ interactions for cases of $(CH_3)_2NH$ to $[(CH_3)_2NH \cdot H_2SO_4]_n$ collisions, and integrated $H_2SO_4$ - $(CH_3)_2NH$ & $H_2SO_4$ - $H_2SO_4$ interactions for cases of $H_2SO_4$ to $[(CH_3)_2NH \cdot H_2SO_4]_n$ collisions based on equation Eq. **8** and obtained the corresponding monomer-cluster potentials: $U_{mc,i}(r) = -\frac{4(n_{c,i}\epsilon_{ii}\sigma_{ii}^6 + n_{c,j}\epsilon_{ij}\sigma_{ij}^6)}{(r^2 - R_c^2)^3}$ and $U_{mc,j}(r) = -\frac{4(n_{c,i}\epsilon_{ji}\sigma_{ji}^6 + n_{c,j}\epsilon_{jj}\sigma_{jj}^6)}{(r^2 - R_c^2)^3}$ where the subscript $i$ and $j$ denote $(CH_3)_2NH$ and $H_2SO_4$ molecules, respectively. Note that based on this way of notation: $\epsilon_{ij} = \epsilon_{ji}$ and $\sigma_{ij} = \sigma_{ji}$. The resulting monomer-cluster interaction potentials were then substituted in Eqs. **3** and **6** to obtain corresponding collision rates. Results are shown in Fig. S5 below.

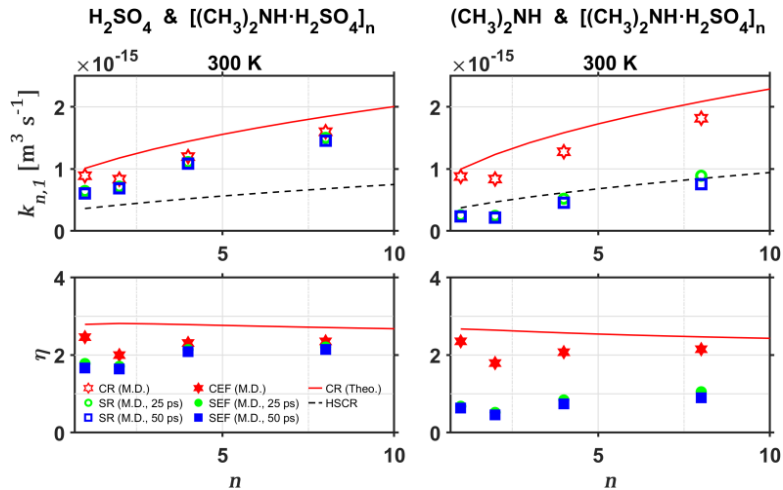

**Figure S5**. Collision rates (CR), sticking rates (SR), hard sphere collision rates (HSCR), collision rate enhancement factor (CEF = CR/HSCR), and sticking rate enhancement factor (SEF = SR/HSCR) for $(CH_3)_2NH$ & $[(CH_3)_2NH \cdot H_2SO_4]_n$ and $H_2SO_4$ & $[(CH_3)_2NH \cdot H_2SO_4]_n$ pairs at 300 K from theoretical predictions (Theo.) and molecular dynamics simulations (M.D. at 25 and 50 ps).